\documentclass{article}
\PassOptionsToPackage{numbers,sort&compress}{natbib}
\usepackage[preprint]{neurips_2025}
\usepackage{amsmath,amssymb}
\usepackage{booktabs}
\usepackage{placeins}
\usepackage{float}
\usepackage{graphicx}
\usepackage[colorlinks=true,linkcolor=blue,citecolor=blue,urlcolor=blue]{hyperref}
\usepackage{xcolor}
\usepackage{tikz}
\usetikzlibrary{arrows.meta,positioning,fit}

\newcommand{\dualitycert}{\textsc{DualityCert}}
\newcommand{\dscode}{\textsc{deepseek-chat}}
\newcommand{\qwcode}{\textsc{qwen-plus}}
\newcommand{\mmcode}{\textsc{MiniMax-M2.5}}

\title{\dualitycert{}: Verifier-Gated Language-Model Repair of Broken
Duality Claims in Quantum Field Theory}
\author{%
  Xingyang Yu\\
  Department of Physics\\
  Virginia Tech\\
  Blacksburg, VA 24061, USA\\
  \texttt{xingyang.jason.yu@gmail.com}
}
\date{\today}

\begin{document}
\maketitle

\begin{abstract}
We present \dualitycert{}, a symbolic verifier for candidate
Seiberg-duality claims in four-dimensional $\mathcal{N}=1$ quiver gauge
theories. The verifier evaluates 't~Hooft anomaly matching,
superpotential $R$-charge consistency, central-charge matching, and a
bounded chiral-ring proxy. A claim that passes receives a consistency
certificate, which states that no tested inconsistency was found, not
that the duality is proven. We use the verifier as a repair environment for
language-model agents, which receive a deliberately broken claim and
must edit it until it certifies. On a preregistered benchmark of 145
broken claims, with the analysis fixed before the first confirmatory
model call, verifier-gated retry improves final repair success over a
single attempt by $+8.3$ percentage points (pp) on \dscode{} and
$+7.1$\,pp on \qwcode{} (Holm-adjusted $p<0.002$). Under an equal
budget of eleven attempts, the stop-first strategy portfolio
underperforms independent verifier-filtered resampling by $10.3$
percentage points on \dscode{} but outperforms it by $14.7$ points on
\qwcode{}, reversing the ordering of the two tested
verifier-exploitation policies across the two confirmatory models. On \qwcode{}, category-level verifier
feedback is worth $+8.7$\,pp over content-free retry, and
interpretable obligation identities alone are worth $+6.4$\,pp over
structurally identical masked feedback. Neither effect is detected on
\dscode{}. Separately, a preregistered \mmcode{} extension again finds
an iteration gain and independent verifier-filtered resampling
outperforming the strategy portfolio. Which policy is better thus
differs between the two models, while every winning policy uses the
same cheap certificate. The verifier, benchmark, protocol, and all
per-attempt records are released.
\end{abstract}

\section{Introduction}
\label{sec:intro}

Large language models are increasingly used to carry out multi-step
derivations in theoretical physics, but their output is usually not
checked by any machine. Judging that output is itself difficult,
because correctness in these domains is layered and often tacit,
so answer matching does not reveal whether the intermediate steps
were reconstructed \cite{Yu:2026ycl}. Benchmarks built from
research-level physics problems find that current models solve few of
them: the research-level tier of a high-energy and cosmology dataset
is mostly unsolved \cite{Chung:2025nsd}, and on research challenges
written by working physicists the best base model averages 5.7
percent \cite{Zhu:2025qnm}. The situation in mathematics is different:
proof assistants provide an exact target for language-model agents to
search against, and systems built on them have progressed from
olympiad-level problem solving
\cite{Yang:2023leandojo,Trinh:2024alphageometry,Hubert:2025alphaproof}
to research-level results such as a machine-verified solution of a
decades-open problem \cite{Sothanaphan:2026erdos728}, together with
formally verified counterexample generation \cite{Li:2026disprove}.

However, this approach
cannot be directly copied to physics. Mathematics is built on an
axiomatic foundation, where a statement is settled by a proof. Formal
quantum field theory (QFT) and string theory, although more rigorous
than most other areas of physics, are largely driven by physical
intuition, and many of their central results are not derived from a
fixed set of axioms. In this regime, where experimental data is
typically unavailable, consistency checks are the main tool by which
physicists judge whether a proposal is correct. Seiberg duality
\cite{Seiberg:1994pq,Intriligator:1995au}, which states that two
different four-dimensional $\mathcal{N}=1$ gauge theories flow to the
same infrared physics, is a representative example: to our knowledge no
proof assistant formalizes this statement,\footnote{The HepLean
project \cite{Tooby-Smith:2024vqu}, since renamed PhysLean and now
part of Physlib, digitalises parts of physics in
Lean~4, including local anomaly cancellation conditions, but not
Seiberg duality.} and the original evidence for
it consists precisely of such consistency checks. A verifier for this
kind of physics should therefore be built on the consistency checks of
the research field itself, rather than as an imitation of a proof
assistant.

A substantial part of these checks
can be made fully mechanical: whether the 't~Hooft anomalies of a
proposed pair of theories match, whether the superpotential (the
holomorphic interaction terms of the theory) is consistent with the
assigned $R$-charges (the charges under the distinguished $U(1)_R$
symmetry), whether the central charges agree, and whether bounded
chiral-ring data agree in a controlled classical approximation. Each
of these is a finite, exact, symbolic computation, and is cheap enough
to run thousands of times.

We present \dualitycert{}, a symbolic verifier based on this layer of
consistency checks. A \emph{candidate duality claim} is an ordered pair
of quiver gauge theories, and \dualitycert{} evaluates four families of
\emph{consistency obligations}: 't~Hooft anomaly matching,
superpotential $R$-charge consistency, central-charge matching, and
$R$-graded bounded chiral-ring consistency. A claim for which at least
one tested obligation passes and none fails receives a
\emph{consistency certificate}. The
certificate states that no tested inconsistency was found. It is not a
proof of duality, and a failed obligation rules out the claim only
within the encoded scope. We then use the verifier to construct a repair
environment: an agent is given a deliberately broken candidate claim,
proposes edits, and each attempt is judged by the verifier. The final
acceptance check is at least as strict as, and for most of the benchmark
stricter than, the checks whose outcomes the agent saw during the
interaction.

We use this environment for a preregistered study of how language-model
agents exploit an exact symbolic verifier. The benchmark consists of 145
repairable claims, obtained by perturbing consistency-certified seed
pairs at depth one in four classes (deleting a superpotential term, flipping a
coefficient sign, perturbing $R$-charges, and changing a gauge-node
rank), and every model receives exactly the same set of claims. The analysis protocol,
the endpoints, the statistical model, and the multiplicity correction
were fixed before the first confirmatory API call. The confirmatory
analysis covers two publicly available models, \dscode{} and \qwcode{}, with
three replications per policy. Effects are estimated by
generalized-estimating-equation (GEE) analysis with fixture-level
clustering under a Holm adjustment and reported as risk differences in
percentage points (pp). Separately, \mmcode{} is evaluated under a
preregistered extension with its own three-hypothesis Holm family
(Section~\ref{sec:results}).

We release \dualitycert{}, the
repair environment, the benchmark, and the preregistered protocol.
On both confirmatory models, verifier-gated iteration improves repair
success over single-shot
attempts ($+8.3$\,pp on \dscode{}, $+7.1$\,pp on
\qwcode{}, Holm-adjusted $p<0.002$). Under an equal budget of
eleven model attempts, the ordering of the two tested exploitation
policies, a stop-first \emph{strategy portfolio} and \emph{independent
verifier-filtered resampling}, reverses between the two named models
($-10.3$\,pp for the portfolio on \dscode{}, $+14.7$\,pp on \qwcode{},
both Holm-significant). The endpoints that isolate the effect of
feedback content show gains only on \qwcode{}, where category-level
verifier feedback is worth
$+8.7$\,pp over content-free retry and interpretable obligation
identities are worth $+6.4$\,pp over structurally identical masked
feedback. Neither gain is detected on \dscode{}. Our claim is
non-universality of the tested policies across
the two evaluated models, not a general classification of models.

Language-model reasoning has been grounded in proof
assistants and symbolic engines. AlphaProof couples an agent to a formal
checker \cite{Hubert:2025alphaproof}, LeanDojo opens the proof-assistant
interaction loop to language models \cite{Yang:2023leandojo}, and
AlphaGeometry pairs a generator with a symbolic deduction engine
\cite{Trinh:2024alphageometry}. In parallel, it is known that unaided
self-correction of model reasoning is unreliable
\cite{Huang:2023selfcorrect}, while external checkers help: learned
verifiers rerank candidate solutions \cite{Cobbe:2021verifiers}, and
execution feedback drives program repair \cite{Chen:2023selfdebug}.
Iterative refinement with feedback is studied in
\cite{Madaan:2023selfrefine,Shinn:2023reflexion}, process-level
verifiers are studied in \cite{Lightman:2023verify}, and program
search filtered by a programmatic evaluator has produced new
mathematical constructions \cite{RomeraParedes:2024funsearch}.
\dualitycert{} differs from these works in two ways: the checker is an
exact, domain-specific symbolic verifier for a QFT duality rather than
a learned reward model, and the object under repair is a physics claim
rather than a program or a formal proof.

Machine-learning classifiers have also been
trained to recognize Seiberg-dual quiver pairs \cite{Bao:2020nbi},
and, in work contemporaneous with this paper, neural networks combined
with pathfinder algorithms learn to trace the mutation sequences that
connect dual quivers \cite{Heckman:2026xsi}. Complementary work on
brane-tiling and coamoeba representations uses unsupervised learning to
identify toric phases and Seiberg-duality phase boundaries
\cite{Seong:2023njx}, conditional generative models to reconstruct brane
configurations and track duality transitions \cite{Seong:2024wkt}, and,
subsequent to this work, explainable convolutional networks to classify
geometric consistency and, within a restricted family, localize defect
fields whose Higgsing restores it \cite{Lee:2026avt}. These approaches
learn to recognize, search for, generate, or diagnose geometric
structures. \dualitycert{} instead evaluates an explicit registry of
field-theoretic obligations exactly within its encoded scope and uses
those checks to gate language-model repair. On the formal side, Lean~4
has recently been used for quantum
field theory. The free bosonic theory has been constructed and shown
to satisfy the Glimm--Jaffe axioms, a variant of the
Osterwalder--Schrader axioms \cite{Douglas:2026hyk},
and the Seiberg--Witten solution of the $\mathcal{N}=2$ $SU(2)$ theory
has been machine-checked from a short list of explicit physical
postulates \cite{Douglas:2026ylk}. Automatic theorem proving has also
been trained on formalized physics \cite{Zhang:2026physprover}. These
projects derive physical results inside proof assistants.
\dualitycert{} addresses a different task. It encodes the consistency
checks that the field already uses as its
evidence, and it runs them cheaply enough to gate the repair loop of
Section~\ref{sec:system}. The nearest methodological comparison is a
multi-agent framework
for quantum many-body simulation that combines a programming verifier
with a physics verifier and corrects errors at each step
\cite{Deng:2026vhj}. That system verifies a research workflow,
and \dualitycert{} verifies a single claim against a fixed obligation
registry, which is what makes the preregistered comparison of
Section~\ref{sec:design} possible.

The rest of the paper is organized as follows. Section~2 describes the
verifier and the repair interface, Section~3 the benchmark and the
preregistered protocol, and Section~4 the confirmatory results. Sections~5
and~6 contain the discussion, limitations, and outlook.

\paragraph{Note added.} The concurrent work of Heckman, Meynet,
Mininno, and Shiu \cite{Heckman:2026xsi} trains neural networks,
assisted by pathfinder algorithms, to search for the mutation
sequences connecting Seiberg-dual quivers. The two papers approach
Seiberg duality from different directions. Theirs is
physics-motivated, and studies the computational complexity of tracing
a duality and how well different architectures learn to do it. Ours is
motivated by AI evaluation, and asks how language-model agents exploit
an exact verifier, with the physics supplying the obligations that
make the verifier exact. The two toolchains are naturally composable:
a learned search can propose candidate duals, and a verifier can
certify them.

\section{\dualitycert{}: certificates and repair interface}
\label{sec:system}

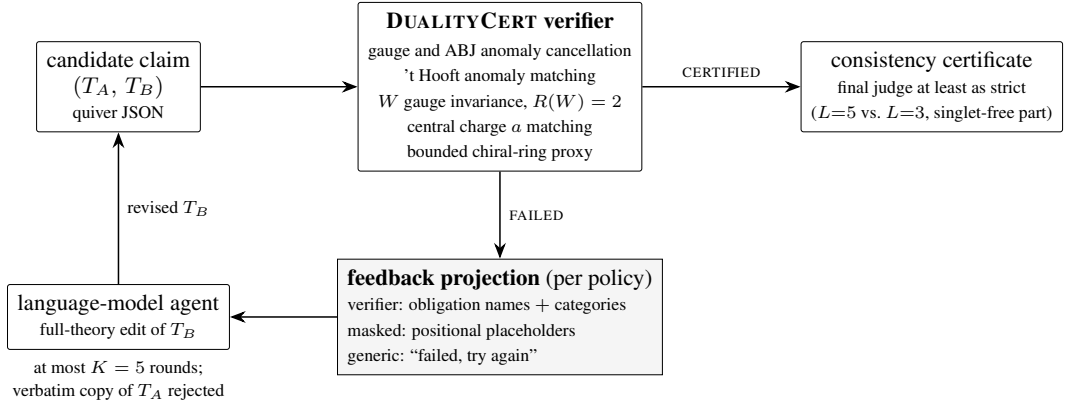
\begin{figure}[t]
\centering
\resizebox{\textwidth}{!}{%
\begin{tikzpicture}[
  font=\footnotesize,
  box/.style={draw, rounded corners=1pt, align=center, inner sep=4pt},
  proj/.style={draw, align=left, inner sep=4pt, fill=black!4},
  arr/.style={-{Stealth[length=2.2mm]}, semithick}]

\node[box, align=center] (claim) at (0,0)
  {candidate claim\\ $(T_A,\,T_B)$\\ \scriptsize quiver JSON};
\node[box, align=center] (ver) at (5.3,0)
  {\textbf{\dualitycert{} verifier}\\[1pt]
   \scriptsize gauge and ABJ anomaly cancellation\\
   \scriptsize 't~Hooft anomaly matching\\
   \scriptsize $W$ gauge invariance, $R(W)=2$\\
   \scriptsize central charge $a$ matching\\
   \scriptsize bounded chiral-ring proxy};
\node[box, align=center] (cert) at (11.3,0)
  {consistency certificate\\
   \scriptsize final judge at least as strict\\
   \scriptsize ($L{=}5$ vs.\ $L{=}3$, singlet-free part)};
\node[box, align=center] (agent) at (0,-3.2)
  {language-model agent\\ \scriptsize full-theory edit of $T_B$};
\node[proj] (fb) at (5.3,-3.2)
  {\textbf{feedback projection} (per policy)\\
   \scriptsize verifier: obligation names $+$ categories\\
   \scriptsize masked: positional placeholders\\
   \scriptsize generic: ``failed, try again''};

\draw[arr] (claim) -- (ver);
\draw[arr] (ver) -- (cert) node[midway, above]{\scriptsize \textsc{certified}};
\draw[arr] (ver) -- (fb) node[midway, right]{\scriptsize \textsc{failed}};
\draw[arr] (fb) -- (agent);
\draw[arr] (agent) -- (claim)
  node[midway, right]{\scriptsize revised $T_B$};
\node[below=1.5mm of agent, inner sep=0pt, align=center]
  {\scriptsize at most $K=5$ rounds;\\[-1pt]
   \scriptsize verbatim copy of $T_A$ rejected};
\end{tikzpicture}}
\caption{The \dualitycert{} repair environment. A candidate claim is
judged by the verifier. Failed obligations pass through a
policy-dependent feedback projection to the agent, which returns a
complete revised $T_B$ (at most $K=5$ rounds; verbatim copies of $T_A$
are rejected). Final acceptance uses a configuration at least as strict
as the interaction-time verifier (Section~\ref{sec:system}).}
\label{fig:overview}
\end{figure}

A candidate duality claim in \dualitycert{} is an ordered pair of
theories $(T_A, T_B)$, conventionally called electric and magnetic.
Figure~\ref{fig:overview} summarizes the construction.
Each theory is specified in a small JSON schema as a quiver gauge
theory: a list of gauge nodes with ranks (labeled
\texttt{SU(2)\_0}, \texttt{SU(2)\_1}, \dots{} in the schema),
a list of arrows (bifundamental or adjoint chiral multiplets)
with their $R$-charges, and a superpotential given as a list of terms,
each a coefficient together with an ordered list of arrow labels. A
theory may also contain gauge-singlet chiral fields, which enter the
superpotential in the same way. The labels electric and magnetic only
name the two encoded descriptions. They are part of the input, not an
outcome of the verification.

\dualitycert{} evaluates the claim against a registry of consistency
obligations. The obligations relevant to this study fall into four
families, and Appendix~\ref{app:verifier} lists the full registry. The
anomaly family requires the cubic gauge anomaly to cancel at every node
of both theories,\footnote{For $SU(2)$ nodes the implemented cubic
condition is a chirality-balance convention rather than a physical
anomaly constraint, since the cubic anomaly of $SU(2)$ vanishes
identically. The $\mathbb{Z}_2$ Witten anomaly \cite{Witten:1982fp} is
not checked. Section~\ref{sec:discussion} returns to this point.} the mixed anomaly
between each gauge node and every encoded global $U(1)$ to cancel in
each theory,\footnote{The benchmark encodes only $U(1)_R$, so this is
the condition that the $R$-symmetry is free of ABJ anomalies, with the
gaugino contribution included.} and the global 't~Hooft anomaly table
of $T_A$ to match that of $T_B$ \cite{tHooft:1979rat}. The
superpotential family requires every superpotential term to be gauge
invariant and to carry $R$-charge two. The $R$-charge family requires
the trial central charge $a$ \cite{Anselmi:1997am}, evaluated on the
encoded $R$-symmetry (no $a$-maximization \cite{Intriligator:2003jj} is
performed), to agree between the two theories.
The chiral-ring family compares bounded classical data: gauge-invariant
single-trace words modulo truncated $F$-term relations, graded by word
length or, when gauge singlets are present, by $R$-charge. This is a
bounded classical proxy for the chiral ring \cite{Cachazo:2002ry}
rather than the ring itself, and Appendix~\ref{app:verifier} states its
exact scope. A claim
for which at least one obligation passes and none fails is
\textsc{certified} and receives a consistency certificate recording the
outcome of every obligation, including those the verifier cannot judge
(\textsc{unknown}, \textsc{not applicable}, or \textsc{not
implemented}). A claim with a failing
in-scope obligation is \textsc{failed}. The certificate therefore
states that no tested obligation failed, not that every conceivable
check was performed, and a failed obligation rules out the claim only
within the encoded scope.

As a concrete example, consider the global 't~Hooft anomaly matching
obligation. For the $U(1)_R$ symmetry, the verifier computes in each
theory the exact rational number
\begin{equation}
\operatorname{Tr} R^3 \;=\; \sum_{a} \bigl(N_a^2-1\bigr)
\;+\; \sum_{X} \dim(\mathcal{R}_X)\,\bigl(R_X-1\bigr)^3 ,
\label{eq:trR3}
\end{equation}
where the first sum runs over the gauge nodes $SU(N_a)$ (the gauginos
carry $R=1$) and the second over the chiral multiplets $X$ in
representation $\mathcal{R}_X$ with $R$-charge $R_X$ (the fermion
component carries $R_X-1$). The obligation holds when the anomaly
tables computed in $T_A$ and $T_B$ agree entry by entry. The linear
trace $\operatorname{Tr} R$ and the mixed global entries are compared in
the same way, while the gauge--global mixed anomalies are separately
required to cancel within each theory. When an entry disagrees, the verifier stores both values
internally, but the repair environment does not show them to the agent:
the feedback reports only the obligation name and its category, in this
case ``global anomaly matching (category: anomaly)''. We deliberately limit the feedback in this way, and we use the
resulting projection as an experimental variable in
Section~\ref{sec:design}.

The environment runs the verifier in two configurations. For fixtures
without gauge singlets (111 of 145), the chiral-ring proxy is evaluated
up to word length $L=3$ (the number of arrow factors in the operator)
during the interaction and up to $L=5$ in the final acceptance check,
so the agent never observes the outcome of the exact check it must
eventually pass. When singlets are present (34 of
145), the automatic $R$-charge grading derives its own cutoffs and the
two configurations coincide. For these fixtures the protection rests on
the withheld numerical residuals and on the copy guard. The copy guard
applies throughout: a candidate that is a verbatim copy of $T_A$ is
rejected by construction, since the identity pair trivially passes all
matching obligations and is not a repair.

The repair interaction itself is simple. In each round the agent
receives $T_A$, the current broken $T_B$, and the feedback from the
previous round, and must return a complete revised $T_B$ in the same
JSON schema (a full-theory edit rather than a patch). Up to $K=5$
rounds are allowed. The content of the feedback depends on the policy:
verifier feedback reports the failed obligation names and categories as
above. Masked feedback keeps the same structure but replaces every
obligation identity by a neutral positional placeholder. Generic retry
states only that verification failed. These feedback projections,
together with the control policies of Section~\ref{sec:design}, define
the five policies of the experiment.

Nothing in this construction is specific to Seiberg duality except the
obligations themselves. The construction uses five reusable
ingredients: a serializable claim schema; a set of consistency
obligations; a certificate recording per-obligation outcomes; a
feedback projection that controls what the agent sees; and a final
judge at least as strict as the interaction-time checks. We regard this as a
design pattern extracted from one implementation, not a validated
general framework. Section~\ref{sec:conclusion} discusses prospects for
building similar verifiers in other areas of formal QFT and string
theory.

\section{Benchmark and preregistered experiment}
\label{sec:design}

We build the benchmark from six seed families of quiver gauge theories
associated with D3-branes probing toric Calabi--Yau threefold
singularities \cite{Douglas:1996sw,Feng:2000mi}, whose quivers and
superpotentials are conveniently encoded by brane tilings
\cite{Hanany:2005ve,Franco:2005rj}: $dP_0$, $dP_1$, $dP_2$, $F_0$, the suspended pinch point
(SPP), and $\mathbb{C}^3/(\mathbb{Z}_2\times\mathbb{Z}_2)$
(Appendix~\ref{app:primer} reviews their origin). For each family we
fix one or more dualized-node choices and gauge ranks (fourteen
seed/node cells in total, with seed ranks between $2$ and $4$, and the
dual ranks then follow from the dualization), and produce a
consistency-certified dual pair by dualizing the chosen node
\cite{Beasley:2001zp}.

We then break a claim by perturbing
the magnetic side $T_B$ at depth one, in one of four classes: deleting
a superpotential term (\texttt{drop\_w\_term}, 38 fixtures), flipping
the sign of one coefficient (\texttt{flip\_w\_sign}, 13), perturbing
$R$-charges (\texttt{r\_charge\_perturb}, 52), or changing one
gauge-node rank (\texttt{rank\_perturb}, 42). We keep a candidate fixture only if the verifier finds that the
perturbed pair fails in scope while the unperturbed
seed pair certifies. This guarantees that a repair is needed and that
at least one certified repair exists. Appendix~\ref{app:benchmark}
records the generation attrition. The resulting 145 repairable fixtures
are generated deterministically from a fixed random-number seed, so the
fixture contents can be regenerated exactly
(Appendix~\ref{app:benchmark} records the provenance-field convention
for the manifest file itself). An execution manifest,
released with the code, records SHA-256 hashes of the generation
configuration and of the fixture list, so the benchmark used in the
experiments can be checked against any regeneration.

We run five policies for every model and refer to them by their
labels in the released artifact. Single-shot repair (\texttt{ss}) makes
one attempt with no feedback. Generic retry (\texttt{gr}) allows up to
$K=5$ rounds and tells the agent only that verification failed.
Verifier feedback (\texttt{vf}) allows the same $K=5$ rounds and
reports the failed obligation names and categories of
Section~\ref{sec:system}. Masked feedback (\texttt{vf\_masked}) is
structurally identical to \texttt{vf} with every obligation identity
replaced by a positional placeholder. The \emph{independent
verifier-filtered resampling} policy (\texttt{best\_of\_n};
best-of-eleven for short) draws up to $2K+1=11$ independent
single-shot attempts, stopping at the first final certificate.

We collect the \texttt{ss}, \texttt{gr}, and \texttt{vf}
components completely for every fixture and replication.
The stop-first \emph{strategy portfolio} (\texttt{ss}, then
\texttt{gr}, then \texttt{vf}, under the same budget of eleven
attempts) is constructed afterwards by a deterministic replay over
these complete components and paired against best-of-eleven, so the
portfolio itself makes no new model calls.

The confirmatory endpoints are pairwise contrasts of final success,
with labels fixed by the preregistered protocol. E2 $=$ \texttt{gr} $-$
\texttt{ss} measures the gain from content-free retry over a single
attempt. E1 $=$ \texttt{vf} $-$ \texttt{gr} estimates the value of the
structured verifier feedback over content-free retry. E4 $=$ portfolio
$-$ \texttt{best\_of\_n} compares the two budget-matched exploitation
policies. E5 $=$ \texttt{vf} $-$ \texttt{vf\_masked} (secondary)
estimates the value of interpretable obligation identities within a
fixed prompt structure. E3 compares first-attempt success under the
\texttt{vf} and \texttt{gr} prompts, written \texttt{vf}@1 and
\texttt{gr}@1 in the tables (descriptive only).

We run the confirmatory campaign on \dscode{} and \qwcode{} with $R=3$
replications per policy, $R$ having been selected in advance from
$\{2,3\}$ by estimating the smallest effect the design could reliably
detect. The analysis protocol fixes the
endpoint definitions, the statistical model, the multiplicity
correction, and the contamination rules. We committed it to the
repository (commit \texttt{813fdb0}) before the first confirmatory
model call and have not modified it since.

Two features of the statistical analysis deserve a plain-language
explanation. Every policy and replication reuses the same 145
fixtures, so outcomes are correlated within a fixture: an intrinsically
hard fixture tends to fail under many policies at once, and the
uncertainty estimates must account for this. We therefore
estimate effects with a binomial generalized-estimating-equation model,
logit link \cite{Liang:1986gee}, policy and replication fixed effects,
fixture as the clustering unit, exchangeable working correlation, and
robust sandwich covariance. The reported effect is a standardized
marginal risk difference with delta-method confidence intervals.
Testing several hypotheses at once also inflates the chance of a
spurious positive, so the primary family $\{$E1, E2,
E4$\}\times\{$\dscode{}, \qwcode{}$\}$ carries one paper-wide Holm
adjustment \cite{Holm:1979seq}, which controls the probability of even
a single false positive across all six tests. E5 forms its own
two-hypothesis family, and E3 is reported with effect and interval
only.

The contamination rules distinguish failures of the
infrastructure from failures of the model. Provider-fault API failures
matching a prespecified predicate (balance-exhaustion responses and
documented provider outages) are treated as exogenous: the affected
fixtures are filtered and re-run under the protocol's resume rule, with
byte-exact quarantine of the replaced records
(Appendix~\ref{app:repro}). Malformed model output is never excluded
and always counts as an outcome. Sampling parameters,
token caps, and the client timeout are listed in
Appendix~\ref{app:repro}.

\section{Confirmatory results}
\label{sec:results}
\FloatBarrier

Table~\ref{tab:primary} collects the preregistered endpoints.
Appendix~\ref{app:stats} lists the per-replication success counts for
the two primary models, and Appendix~\ref{app:minimax} lists those for
the extension. Appendix~\ref{app:repro} records the compute budget of
the primary campaigns. The table and the figure of this
section are generated directly from the released run artifacts by a
single script, and we take every number quoted below from
them.

\begin{table}[H]
\centering
\begin{tabular}{llrrr}
\toprule
model & endpoint & RD (pp) & 95\% CI & $p_{\mathrm{Holm}}$ \\
\midrule
\textsc{deepseek-chat} & E2 (\texttt{gr} $-$ \texttt{ss}) & +8.3 & [+3.8, +12.8] & 0.0015 \\
\textsc{deepseek-chat} & E1 (\texttt{vf} $-$ \texttt{gr}) & -1.8 & [-6.6, +3.0] & 0.4528 \\
\textsc{deepseek-chat} & E4 (portfolio $-$ \texttt{best\_of\_n}) & -10.3 & [-16.2, -4.5] & 0.0016 \\
\textsc{qwen-plus} & E2 (\texttt{gr} $-$ \texttt{ss}) & +7.1 & [+3.2, +11.1] & 0.0016 \\
\textsc{qwen-plus} & E1 (\texttt{vf} $-$ \texttt{gr}) & +8.7 & [+3.8, +13.7] & 0.0016 \\
\textsc{qwen-plus} & E4 (portfolio $-$ \texttt{best\_of\_n}) & +14.7 & [+8.4, +21.0] & $<10^{-4}$ \\
\midrule
\textsc{deepseek-chat} & E5 (\texttt{vf} $-$ \texttt{vf\_masked}) & +0.0 & [-4.6, +4.6] & 1.0000 \\
\textsc{qwen-plus} & E5 (\texttt{vf} $-$ \texttt{vf\_masked}) & +6.4 & [+2.2, +10.7] & 0.0065 \\
\textsc{deepseek-chat} & E3 (\texttt{vf}@1 $-$ \texttt{gr}@1) & +0.5 & [-3.4, +4.3] & -- \\
\textsc{qwen-plus} & E3 (\texttt{vf}@1 $-$ \texttt{gr}@1) & +1.8 & [-1.1, +4.8] & -- \\
\midrule
\multicolumn{5}{l}{\emph{preregistered \textsc{MiniMax-M2.5} extension (separate three-hypothesis Holm family)}} \\
\midrule
\textsc{MiniMax-M2.5} & E2 (\texttt{gr} $-$ \texttt{ss}) & +11.5 & [+5.9, +17.1] & 0.0002 \\
\textsc{MiniMax-M2.5} & E1 (\texttt{vf} $-$ \texttt{gr}) & +2.8 & [-2.0, +7.5] & 0.2590 \\
\textsc{MiniMax-M2.5} & E4 (portfolio $-$ \texttt{best\_of\_n}) & -9.0 & [-15.2, -2.7] & 0.0094 \\
\textsc{MiniMax-M2.5} & E5 (\texttt{vf} $-$ \texttt{vf\_masked}) & -0.7 & [-5.3, +3.9] & 0.770$^\dagger$ \\
\bottomrule
\end{tabular}

\caption{Preregistered GEE endpoints: standardized marginal risk differences
in percentage points, with fixture-clustered robust 95\% CIs. Rows
follow the ladder order of Section~\ref{sec:design}. The upper block
reports the primary two-model campaign: its first six rows form the
primary family with one paper-wide Holm adjustment, E5 forms a
separate two-hypothesis family, and E3 is descriptive. The lower block
is the separately preregistered \mmcode{} extension with its own
three-hypothesis Holm family. $\dagger$: unadjusted $p$ (the extension
E5 is secondary). E3 is descriptive and reported with effect and
interval only.}
\label{tab:primary}
\end{table}

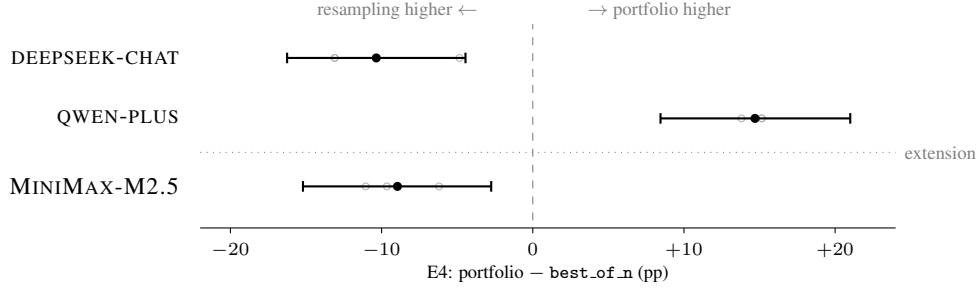
\begin{figure}[t]
\centering
\begin{tikzpicture}[font=\footnotesize]
\draw[dashed, gray] (4.40,-0.15) -- (4.40,2.45);
\draw (0.00,-0.25) -- (9.20,-0.25);
\draw (0.40,-0.25) -- (0.40,-0.32) node[below] {\scriptsize $-20$};
\draw (2.40,-0.25) -- (2.40,-0.32) node[below] {\scriptsize $-10$};
\draw (4.40,-0.25) -- (4.40,-0.32) node[below] {\scriptsize $0$};
\draw (6.40,-0.25) -- (6.40,-0.32) node[below] {\scriptsize $+10$};
\draw (8.40,-0.25) -- (8.40,-0.32) node[below] {\scriptsize $+20$};
\node[below] at (4.60,-0.62) {\scriptsize E4: portfolio $-$ \texttt{best\_of\_n} (pp)};
\node[anchor=east, gray] at (3.80,2.62) {\scriptsize resampling higher $\leftarrow$};
\node[anchor=west, gray] at (5.00,2.62) {\scriptsize $\rightarrow$ portfolio higher};
\draw[dotted, gray] (0.00,0.75) -- (9.20,0.75) node[pos=1, right] {\scriptsize extension};
\node[anchor=east] at (-0.15,2.0) {\textsc{deepseek-chat}};
\draw[gray!60] (1.78,2.0) circle (0.045);
\draw[gray!60] (1.78,2.0) circle (0.045);
\draw[gray!60] (3.43,2.0) circle (0.045);
\draw[thick] (1.15,2.0) -- (3.51,2.0);
\draw[thick] (1.15,1.93) -- (1.15,2.07);
\draw[thick] (3.51,1.93) -- (3.51,2.07);
\fill (2.33,2.0) circle (0.06);
\node[anchor=east] at (-0.15,1.2) {\textsc{qwen-plus}};
\draw[gray!60] (7.43,1.2) circle (0.045);
\draw[gray!60] (7.16,1.2) circle (0.045);
\draw[gray!60] (7.43,1.2) circle (0.045);
\draw[thick] (6.09,1.2) -- (8.60,1.2);
\draw[thick] (6.09,1.13) -- (6.09,1.27);
\draw[thick] (8.60,1.13) -- (8.60,1.27);
\fill (7.34,1.2) circle (0.06);
\node[anchor=east] at (-0.15,0.3) {\textsc{MiniMax-M2.5}};
\draw[gray!60] (2.19,0.3) circle (0.045);
\draw[gray!60] (2.47,0.3) circle (0.045);
\draw[gray!60] (3.16,0.3) circle (0.045);
\draw[thick] (1.36,0.3) -- (3.85,0.3);
\draw[thick] (1.36,0.22999999999999998) -- (1.36,0.37);
\draw[thick] (3.85,0.22999999999999998) -- (3.85,0.37);
\fill (2.61,0.3) circle (0.06);
\end{tikzpicture}
\caption{The E4 contrast: success-rate difference between the
stop-first strategy portfolio and independent verifier-filtered
resampling (\texttt{best\_of\_n}) under the same budget of eleven
attempts. Filled points are
the GEE estimates, horizontal bars the fixture-clustered 95\% CIs, and
small open circles the three per-replication differences. Negative
values favor resampling, positive values favor the portfolio. The
separately preregistered extension model appears below the dotted
divider.}
\label{fig:e4forest}
\end{figure}

For E2, content-free retry improves final repair
success over a single attempt by $+8.3$\,pp on \dscode{} (Holm
$p=0.0015$) and $+7.1$\,pp on \qwcode{} ($p=0.0016$). The sign is the
same in every replication of both models, which supports a stable
iteration gain on this benchmark.

E4 changes sign between the two models, as shown in
Figure~\ref{fig:e4forest}. Under
the same budget of eleven attempts, the strategy portfolio loses to
independent verifier-filtered resampling by $10.3$\,pp on \dscode{}
($p=0.0016$) and beats it by $14.7$\,pp on \qwcode{} ($p<10^{-4}$),
again with consistent per-replication signs on both models. The
ordering of the two tested exploitation policies therefore reverses
between the two named models. On \dscode{} the better use of the
eleven attempts is to sample independently and let the verifier
filter, and on \qwcode{} it is to escalate through the feedback
policies.

E1 and E5 locate the difference in the feedback channel.
On \qwcode{}, structured verifier feedback is worth $+8.7$\,pp over
content-free retry (E1, $p=0.0016$), and interpretable obligation
identities alone are worth $+6.4$\,pp over structurally identical
masked feedback (E5, $p=0.0065$). On \dscode{} we detect neither
effect: E1 is $-1.8$\,pp ($p=0.45$) and E5 is $+0.0$\,pp ($p=1.0$).
The first-attempt contrast E3 is small on both models ($+0.5$ and
$+1.8$\,pp, descriptive), which is consistent with the gain from
feedback content arriving in the later rounds rather than at the first
attempt.

For the \mmcode{} extension (lower block of
Table~\ref{tab:primary}, below the divider in
Figure~\ref{fig:e4forest}), we analyze the three preregistered
endpoints under the extension's own three-hypothesis Holm family. The extension reproduces the
E2 anchor, $+11.5$\,pp
($p=2\times10^{-4}$), and it reproduces the negative E4 ordering
observed on \dscode{}: the portfolio loses to resampling by
$9.0$\,pp ($p=0.0094$), with consistent per-replication signs. The
feedback endpoints are not significant on this model (E1 $+2.8$\,pp;
E5 $-0.7$\,pp, secondary). Provider-forced reasoning, elevated invalid
rates, and the transport-error cleaning behind these numbers are
discussed in Section~\ref{sec:discussion} and
Appendix~\ref{app:minimax}.

The verifier's cheap consistency certificate supports the
higher-performing E4 policy in each campaign, either as
a filter over independent samples or as a source of
feedback. What changes between the named models is which
exploitation policy wins. The data support non-universality of the
tested policies across the two evaluated models, and they do not
support a general classification of models.

\section{Discussion and limitations}
\label{sec:discussion}

E4 compares the strategy portfolio with independent verifier-filtered
resampling under the same budget of eleven attempts. These two
policies differ in several ways at once: the portfolio's later
attempts receive feedback, its attempts depend on earlier ones, and
its budget is spread over three stages rather than eleven independent
draws. The sign change between the two models in
Section~\ref{sec:results} therefore does not, by itself, identify
which of these ingredients is responsible. E1 and E5 give partial information. In particular, feedback content
helps on
\qwcode{} and shows no detectable effect on \dscode{}. However, this
does not identify a single mechanism behind the sign change. All of our claims concern the tested policies on the named
models.

E5 estimates the value of interpretable obligation identities beyond
failure count and list structure. The preregistered protocol document, released with the code, states
this limitation before any data was collected, and we repeat it
verbatim: ``neutral opaque placeholders
may themselves change model attention and epistemic behavior; if that
effect is large, E5 still will not isolate semantic information despite
avoiding outcome-dependent misinformation.'' The interpretability
premium of $+6.4$\,pp is therefore a finding about \qwcode{} under this
specific construction, not a general property of verifier feedback.

The extension model differs from the primary campaigns in several
ways. The provider forces its
reasoning mode on, so its attempts are generated differently from the
primary models. Invalid final outcomes are frequent for every model in
this study (on the \texttt{vf} policy, 35\% for \dscode{}, 32\% for
\qwcode{}, and 45\% for \mmcode{}, with \texttt{gr} between 34\% and
44\%), and the extension's \texttt{vf} rate is the highest. In
addition, some attempts of the extension campaign failed for reasons
unrelated to the model (dropped connections and timeouts). The
affected records were filtered and re-run, and
Appendix~\ref{app:minimax} documents the audited procedure, the exact
counts, and the preserved original records.

Two models support an existence claim of non-universality,
not a distribution over models. The domain is one
duality class in four-dimensional $\mathcal{N}=1$ quiver gauge
theories. The certificate is a consistency statement whose reach is set
by the encoded obligations, and Section~\ref{sec:system} records
where the encoded scope is narrower than the physics. The $SU(2)$ cubic
condition is a chirality-balance convention and the Witten anomaly
\cite{Witten:1982fp} is not checked, and the chiral-ring comparison is a bounded classical
proxy whose feedback and final gradings coincide on singlet-containing
fixtures. In addition, the $R$-charge obligations test the encoded
assignment for consistency but do not perform the $a$-maximization
\cite{Intriligator:2003jj} that determines the superconformal $R$-charges at the infrared fixed
point, even though the seed theories flow to such fixed points, so the
central charges compared are trial values. At depth two, exploratory \dscode{} campaigns find every tested
policy near the floor, with success at or below 7\% on 120 fixtures
in single replications (Appendix~\ref{app:depth2}). Harder instances
therefore remain well beyond all policies tested here.

\section{Conclusion, artifact statement, and outlook}
\label{sec:conclusion}

\dualitycert{} turns a set of consistency checks for Seiberg duality
in quantum field theory into a machine-checkable certificate, and the
certificate into a repair environment for language-model agents. On the preregistered
benchmark, retry gated by the verifier reliably improves repair
success on both primary models, while the better of the two
budget-matched exploitation policies differs between the two named
models, and the cheap certificate is what every winning policy uses.
All components are released with this paper: the verifier and the
repair environment,
the benchmark with its generation configuration and manifest hashes,
the preregistered protocol and its amendment, the per-attempt records
of every confirmatory campaign and of the exploratory depth-two
campaigns, the contamination audit trail, and the commands
that regenerate the tables and the results figure of this paper from
those artifacts. The release is available at
\url{https://github.com/xingyang-yu/QFTCert} under the Apache License 2.0.

Nothing in the construction of Section~\ref{sec:system} is specific to
Seiberg duality except the obligations themselves. We view
\dualitycert{} as the first step of a prospective broader program,
which we call \textsc{QFTCert}: verifiers built on the native
consistency checks of other areas of formal QFT and string theory,
following the same pattern of claim schema, obligation set,
certificate, feedback projection, and strict final judge. The nearest
extensions are other dualities in quantum field theory, for instance
triality in two dimensions, mirror symmetry in three dimensions, and
S-duality in four dimensions. Beyond dualities, natural candidates
include crossing symmetry and unitarity in the conformal bootstrap,
factorization and soft limits for scattering amplitudes, exact
protected quantities of superconformal field theories (and even the
classification of supersymmetric field theories),
tadpole and anomaly cancellation in the string landscape, the fusion
and anomaly data of generalized global symmetries, entanglement
inequalities and causality constraints in bottom-up holography and
quantum gravity, and consistency conditions in string
field theory.\footnote{This is by no means the full list of potential
directions, and not even a small portion of it. These are simply the
directions the author is familiar with and has some command of.}
Any area that maintains sharp consistency checks of its own is a
candidate. Encoding an area
requires its own domain expertise, and the pattern is designed so that
areas can be added independently. We would welcome
such extensions, and we hope that formal theoretical physics gradually
acquires machine-checkable substrates of this kind, on which
language-model agents can be evaluated, compared, and eventually
used for substantial derivational work.

\begin{ack}
This work was supported by NSF grant PHY-2310588. The author thanks
Jonathan J. Heckman for fruitful discussions, and Anthropic for providing
access to Claude Max plan during the early development of this work.
\end{ack}

\bibliographystyle{unsrtnat}
\bibliography{refs}

\appendix

\section{Physics primer: Seiberg duality and the consistency obligations}
\label{app:primer}

Seiberg duality \cite{Seiberg:1994pq} is the statement that two
different four-dimensional $\mathcal{N}=1$ gauge theories can flow to
the same infrared physics. In the original example the electric theory
is $SU(N_c)$ QCD with $N_f$ flavors, and the magnetic theory is
$SU(N_f-N_c)$ with $N_f$ flavors, gauge-singlet mesons, and a cubic
superpotential. The two descriptions can be weakly or strongly coupled
in different regimes, which makes the proposal powerful but also hard
to test directly. The original evidence consists of consistency
checks: matching of 't~Hooft anomalies, matching of gauge-invariant
chiral operators and of moduli spaces of vacua, and consistent
behavior under mass deformations
\cite{Seiberg:1994pq,Intriligator:1995au}.

The theories in this paper are quiver gauge theories
\cite{Douglas:1996sw}. A quiver is a directed graph. Each node carries a gauge factor $SU(N_a)$, each arrow
from node $a$ to node $b$ is a chiral multiplet in the bifundamental
representation $(N_a,\bar N_b)$, an arrow from a node to itself is an
adjoint, and gauge-singlet chiral fields may also be present. The
interactions are encoded in a superpotential $W$, a holomorphic
polynomial in the fields, here built from closed cycles of the quiver.
Each chiral field carries a charge under the global $U(1)_R$ symmetry,
its $R$-charge, and the simplest gauge-invariant chiral operators are
built from closed paths and singlets.

The logic of the verifier rests on necessary conditions. If two
theories really flow to the same infrared physics, every quantity
protected along the renormalization-group flow must agree between
them. A disagreement in a protected quantity therefore rules out the
proposed equivalence, within the scope of what is encoded. Agreement
in finitely many protected quantities, however, is never a proof. It
only says that the proposal survived the tests it was given, and this
is the precise content of a consistency certificate.

The sharpest such protected quantities are the 't~Hooft anomalies of
global symmetries \cite{tHooft:1979rat}. Their coefficients are
one-loop exact and invariant along the flow, so they must agree
between dual descriptions. The anomalies can be computed in either
description at any scale, and if both flows end at the same
infrared physics the results have to coincide. For the $U(1)_R$ symmetry the relevant
traces are $\operatorname{Tr}R$ and $\operatorname{Tr}R^3$ over the
fermions, computed as in Eq.~\eqref{eq:trR3}, with gauginos entering
at $R=1$. The same traces determine the conformal central charge
\cite{Anselmi:1997am},
\begin{equation}
a \;=\; \frac{3}{32}\left(3\operatorname{Tr}R^3
- \operatorname{Tr}R\right) ,
\label{eq:centralcharge}
\end{equation}
which gives the physical central charge of the infrared fixed point
only when evaluated on the exact superconformal $R$-symmetry. \dualitycert{} evaluates these quantities on
the $R$-charges encoded in the claim. It checks the assignment, and it
does not derive it.

A consistent $R$-assignment must satisfy two further conditions. Every
superpotential term must carry $R$-charge exactly two, because $W$
enters the action through an integral over half of superspace. The
$R$-symmetry must also be free of mixed anomalies with every gauge
factor, with the gaugino contribution included, or it is not a
symmetry of the quantum theory. At an infrared fixed point the exact
superconformal $R$-symmetry is singled out among all consistent
assignments by $a$-maximization \cite{Intriligator:2003jj}. The
verifier does not perform this maximization. It tests the encoded
assignment against the two conditions above.

The classical chiral ring consists of the gauge-invariant chiral
operators modulo the $F$-term relations of the superpotential, and
quantum effects can deform these relations \cite{Cachazo:2002ry}.
Dual descriptions of the same infrared physics should agree on this
protected data. A full comparison would involve baryons, products of
traces, finite-$N$ trace identities, and quantum-corrected relations.
\dualitycert{} instead compares a bounded classical piece:
single-trace cyclic words, that is closed arrow paths counted up to
cyclic rotation, modulo truncated $F$-term relations, counted
by word length up to a cutoff, or by $R$-charge when gauge singlets
are present. Word length is not invariant across a duality (a
singlet meson on one side maps to a two-arrow composite on the
other), so this comparison is a proxy within the encoded scope rather
than a physical necessary condition. The roles of the cutoffs $L=3$
and $L=5$ are described in Section~\ref{sec:system}.

The seed theories of the benchmark describe D3-branes probing toric
Calabi--Yau threefold singularities \cite{Douglas:1996sw,Feng:2000mi},
with quivers and superpotentials conveniently encoded by brane tilings
\cite{Hanany:2005ve,Franco:2005rj}. Brane tilings are also subject to
geometric consistency conditions with combinatorial formulations
\cite{Hanany:2015tgh}. This notion is distinct from, and is not currently
included among, the field-theoretic obligations checked by
\dualitycert{}. Here $dP_0$, $dP_1$, and $dP_2$
denote the complex cones over del Pezzo surfaces ($\mathbb{P}^2$ blown
up at zero, one, and two points), $F_0$ the cone over
$\mathbb{P}^1\times\mathbb{P}^1$, SPP the suspended pinch point, and
$\mathbb{C}^3/(\mathbb{Z}_2\times\mathbb{Z}_2)$ an abelian orbifold of
flat space. At suitable nodes, Seiberg duality acts on a single gauge factor of
such a quiver and relates different toric phases of the same
singularity \cite{Beasley:2001zp}. The benchmark construction on top of these
seeds is described in Appendix~\ref{app:benchmark}.

\FloatBarrier
\section{Verifier, benchmark, and interface details}
\label{app:system-detail}

\subsection{The verifier}
\label{app:verifier}
\FloatBarrier

The verifier evaluates a registry of twenty-three obligations under
the profile used throughout this paper. Table~\ref{tab:registry}
lists every obligation together with the status it receives on a
committed positive fixture: eleven are judged and certified, and the
remaining twelve are recorded without being judged under the default
policy. The four families of Section~\ref{sec:system} correspond to
the anomaly, superpotential, central-charge, and chiral-ring entries
of the table.

\begin{table}[H]
\centering
\begin{tabular}{ll}
\toprule
obligation & status on a positive fixture \\
\midrule
theory kind classification & \textsc{certified} \\
electric gauge anomaly cancellation & \textsc{certified} \\
magnetic gauge anomaly cancellation & \textsc{certified} \\
electric gauge-global mixed anomaly cancellation & \textsc{certified} \\
magnetic gauge-global mixed anomaly cancellation & \textsc{certified} \\
electric superpotential consistency & \textsc{certified} \\
magnetic superpotential consistency & \textsc{certified} \\
global symmetry matching & \textsc{certified} \\
global anomaly matching & \textsc{certified} \\
central charge matching from encoded R-symmetry & \textsc{certified} \\
bounded chiral-ring consistency & \textsc{certified} \\
\midrule
superconformal R-charge audit & \textsc{not applicable} \\
a-maximization central charge matching & \textsc{not applicable} \\
SCFT soundness (necessary conditions) & \textsc{not applicable} \\
index matching & \textsc{not applicable} \\
\midrule
operator map Abelian charge matching & \textsc{not implemented} \\
deformation checks & \textsc{not implemented} \\
\midrule
operator unitarity bound from encoded R-symmetry & \textsc{unknown} \\
chiral ring / F-term metadata & \textsc{unknown} \\
moduli-space metadata & \textsc{unknown} \\
conformal-manifold metadata & \textsc{unknown} \\
generalized-symmetry / defect metadata & \textsc{unknown} \\
protected quantity hooks & \textsc{unknown} \\
\bottomrule
\end{tabular}

\caption{The full obligation registry under the paper's verifier
profile, grouped by the status each obligation receives on a committed
positive fixture. The table is generated by running the verifier, so
it always reflects the released code.}
\label{tab:registry}
\end{table}

Each obligation receives one of five statuses: \textsc{certified},
\textsc{failed}, \textsc{unknown}, \textsc{not applicable}, or
\textsc{not implemented}. A separate \textsc{error}
status exists for the verification outcome as a whole, for malformed
inputs, and routes the pair to attrition. A claim is in scope when its
overall verdict is \textsc{certified} or \textsc{failed}.
Certification requires that at
least one obligation is \textsc{certified} and none fails, and
obligations the verifier cannot judge are recorded in the certificate
without blocking certification, as discussed in
Section~\ref{sec:system}. In Table~\ref{tab:registry}, the
\textsc{not applicable} rows are opt-in or profile-gated checks. For
instance, index matching computes the superconformal indices of the
two theories as bounded series, but runs only when explicitly enabled
in the claim metadata. The \textsc{unknown} rows compare optional
encoded data, such as externally computed protected quantities, and
no benchmark fixture encodes such data. The certificate records the status and
message of every obligation together with detailed numerical tables,
for example the full anomaly tables of both sides.

The chiral-ring comparison enumerates single-trace cyclic words
(closed arrow paths counted up to cyclic rotation) modulo the
truncated $F$-term ideal. In word-length mode the interaction-time
verifier uses $L=3$ and the final judge $L=5$. When either side
contains gauge singlets, the comparison switches automatically to
$R$-charge grading with maximal $R$-charge $2$, and the per-side depth
cutoffs are then derived from the $R$-charge range, which makes the
two configurations coincide, as disclosed in
Section~\ref{sec:system}.

\FloatBarrier
\subsection{The benchmark}
\label{app:benchmark}
\FloatBarrier

Table~\ref{tab:seeds} lists the seed catalog: the six geometries of
Appendix~\ref{app:primer} with their artifact labels, seed ranks,
dualized nodes, and fixture counts, fourteen family/rank/node cells in
total. For each cell the electric theory comes from the seed catalog,
the magnetic side is produced by dualizing the chosen node with the
package's mutation engine, and the pair is kept only if it certifies.

\begin{table}[H]
\centering
\begin{tabular}{llccc}
\toprule
geometry & artifact label & seed ranks $N$ & dualized nodes & fixtures \\
\midrule
$dP_0$ (cone over $\mathbb{P}^2$) & \texttt{dp0\_toric} & 3, 4 & 0, 1, 2 & 49 \\
$dP_1$ & \texttt{dp1} & 2 & 0, 1 & 19 \\
$dP_2$ & \texttt{dp2\_phase1} & 2 & 3, 4 & 17 \\
$F_0$ ($\mathbb{P}^1\times\mathbb{P}^1$) & \texttt{f0\_phase\_ii} & 3 & 0, 2 & 26 \\
SPP & \texttt{spp} & 2 & 0, 2 & 18 \\
$\mathbb{C}^3/(\mathbb{Z}_2\times\mathbb{Z}_2)$ & \texttt{c3\_z2z2} & 2 & 0, 1 & 16 \\
\midrule
total & 14 family/rank/node cells & & & 145 \\
\bottomrule
\end{tabular}

\caption{The seed catalog behind the benchmark. Geometry names follow
Appendix~\ref{app:primer}, artifact labels follow the released code,
and the dualized-node column lists the nodes used across the seed
ranks of that family.}
\label{tab:seeds}
\end{table}

The four perturbation operators act on the magnetic side and record
their action in the manifest:
\begin{itemize}
\item \texttt{drop\_w\_term} removes one superpotential term;
\item \texttt{flip\_w\_sign} negates one coefficient;
\item \texttt{r\_charge\_perturb} shifts the $R$-charge of one
      field by a recorded rational amount (for example $-1/3$);
\item \texttt{rank\_perturb} changes one gauge rank by one unit.
\end{itemize} A
candidate fixture is kept only if the perturbed pair fails in scope
while the seed pair certifies. Of the generated candidates, 79 were
dropped and recorded: 27 duplicates and 52 whose verifier label did
not match the expected one. The copy-cheat guard operates at repair
time: a revised $T_B$ that is a verbatim copy of $T_A$ is rejected,
since the identity pair trivially certifies. The fixture content regenerates deterministically from the seed in
the configuration file, with the commands of
Appendix~\ref{app:repro}. The manifest additionally records
generation provenance (a timestamp and a code commit), so reproducing
the manifest file itself byte for byte requires fixing these
provenance fields.

\FloatBarrier
\subsection{Prompts and policies}
\label{app:prompts}

This subsection records the exact agent interface for
auditability, as reference material.

The agent is addressed by a fixed system prompt that begins ``You are a theoretical physicist repairing a proposed 4d N=1
supersymmetric gauge theory duality''
and receives Theory A, the current Theory B, and the feedback text of
the previous round. The reply is forced through a JSON tool schema
whose fields are an action (\texttt{edit\_candidate},
\texttt{no\_change}, or \texttt{abstain}), the complete revised
theory for an edit, and a reasoning field capped at 300 characters by the schema and
instructed to be one short sentence. The full prompt strings and schemas ship with the code.

The feedback texts are fixed strings. Generic retry always receives
``The candidate failed verification. Try a different edit to make
Theory B a valid dual of Theory A.'' Verifier feedback receives the
preamble ``The candidate failed verification. Failed obligations:''
followed by one bullet per failed obligation, each of the form
``\texttt{<name> (category: <category>)}''. Masked feedback receives
the same preamble and bullet structure with every identity replaced by
``\texttt{obligation-$i$ (category: category-$i$)}''. A candidate that already passes the interaction-time verifier is
recorded as already consistent, without a model call.

A round is invalid when the model returns no parsable action or when
the action fails to apply, for example a malformed theory. An
abstention is an explicit \texttt{abstain} action. For the feedback
policies, the first invalid round or abstention ends the fixture as
unrepaired, and the fixture-level invalid flag records an invalid
round. \texttt{best\_of\_n} instead continues past invalid and
abstaining draws, and its fixture-level invalid flag is set only when
no draw reached verification and at least one draw was invalid.

\FloatBarrier
\section{Statistical specification}
\label{app:stats}

This appendix records the preregistered analysis at estimator
level. It is reference material for checking the reported
numbers, and the plain-language summary is in
Section~\ref{sec:design}.

\begin{table}[H]
\centering
\begin{tabular}{llccc}
\toprule
model & policy & rep 1 & rep 2 & rep 3 \\
\midrule
\textsc{deepseek-chat} & \texttt{ss} & 8/145 & 11/145 & 17/145 \\
 & \texttt{gr} & 22/145 & 24/145 & 26/145 \\
 & \texttt{vf} & 20/145 & 25/145 & 19/145 \\
 & \texttt{vf\_masked} & 15/145 & 25/145 & 24/145 \\
 & \texttt{best\_of\_n} & 58/145 & 65/145 & 59/145 \\
\midrule
\textsc{qwen-plus} & \texttt{ss} & 10/145 & 11/145 & 9/145 \\
 & \texttt{gr} & 19/145 & 18/145 & 24/145 \\
 & \texttt{vf} & 28/145 & 32/145 & 39/145 \\
 & \texttt{vf\_masked} & 13/145 & 30/145 & 28/145 \\
 & \texttt{best\_of\_n} & 21/145 & 26/145 & 27/145 \\
\bottomrule
\end{tabular}

\caption{Per-replication verifier-certified repair counts (out of 145)
for every policy, model, and replication of the primary campaign.}
\label{tab:perrep}
\end{table}

For each model and endpoint the protocol specifies a marginal binomial
GEE fit with logit link, policy and
replication fixed effects, fixture identity as the clustering unit,
exchangeable working correlation, and robust sandwich covariance. The
reported effect is a standardized marginal risk difference: each policy
is predicted at every observed replication level, predictions are
averaged equally over replication levels and the 145 fixtures, and the
two policy averages are subtracted. Confidence intervals come from the
delta method on the robust covariance. If the exchangeable fit fails
numerically, the same mean model is refit with independence working
correlation and the same robust covariance, and this fallback is
triggered only by a logged solver failure. Identity-link fits and
fixture bootstraps are excluded by the protocol.

The primary family consists of the six hypotheses \{E1, E2,
E4\}$\times$\{\dscode{}, \qwcode{}\} under one Holm adjustment: the
six $p$-values are sorted, the smallest is compared against $0.05/6$,
the next against $0.05/5$, and so on, stopping at the first failure.
E5 forms its own two-hypothesis family, and E3 is reported with effect
and interval only (Table~\ref{tab:primary}).

The replication count $R=3$ was selected from $\{2,3\}$ before the
protocol freeze, using a prospective minimum-detectable-effect table
and the observed cost per replication. For success-probability
variance bounds $q\in\{0.10, 0.15, 0.20\}$, at two-sided level
$0.05/6$ and $80\%$ power, the detectable risk differences are
$6.5$, $7.9$, and $9.1$\,pp at $R=2$, and $5.3$, $6.5$, and
$7.5$\,pp at $R=3$. The protocol labels these values an optimistic
independence bound: positive within-fixture correlation $\rho$
inflates them by $\sqrt{1+(R-1)\rho}$.

\FloatBarrier
\section{Additional descriptive results}
\label{app:descriptive}

\subsection{Per-class outcomes}
\label{app:perclass}

Table~\ref{tab:perclass} lists verifier-certified repair counts per
perturbation class for the \texttt{gr} and \texttt{vf} policies,
summed over the three replications (out of three times the class
size). These counts are descriptive. The per-class pattern varies
between models and classes, and no per-class hypothesis was
preregistered, so we draw no inference from this table.

\begin{table}[H]
\centering
\begin{tabular}{lcccc}
\toprule
 & \multicolumn{2}{c}{\textsc{deepseek-chat}} & \multicolumn{2}{c}{\textsc{qwen-plus}} \\
class & \texttt{gr} & \texttt{vf} & \texttt{gr} & \texttt{vf} \\
\midrule
\texttt{drop\_w\_term} & 9/114 & 12/114 & 14/114 & 21/114 \\
\texttt{flip\_w\_sign} & 7/39 & 4/39 & 8/39 & 10/39 \\
\texttt{r\_charge\_perturb} & 32/156 & 18/156 & 25/156 & 33/156 \\
\texttt{rank\_perturb} & 24/126 & 30/126 & 14/126 & 35/126 \\
\bottomrule
\end{tabular}

\caption{Per-class verifier-certified repair counts of the two primary
models for content-free
retry (\texttt{gr}) and verifier feedback (\texttt{vf}), summed over
three replications. Descriptive only.}
\label{tab:perclass}
\end{table}

\FloatBarrier
\subsection{Depth-two and excluded runs}
\label{app:depth2}

Two exploratory \dscode{} campaigns were run on 120 fixtures built
with two perturbations instead of one, each at one replication per
policy. The first campaign gave verifier-certified repairs on 7/120
fixtures for single-shot, 8/120 for content-free retry, 5/120 for
category-level feedback, and 2/120 for a detailed-feedback variant.
The second campaign covered the first three policies and gave 0/120,
3/120, and 1/120, with a higher rate of invalid final outcomes. An
attempt between the two campaigns failed on provider-side balance
errors before producing any model output. All three run sets are
retained in the released artifacts. All tested policies sit near the
floor in both campaigns, the ordering carries no inferential weight
at one replication, and no depth-two result enters any confirmatory
claim. The depth-two fixtures are therefore substantially harder than
the depth-one fixtures used in the confirmatory experiment.

Two early scouting runs were excluded before the confirmatory phase
for infrastructure reasons. A free-tier deployment of an unrelated
model produced heavy provider-side throttling noise, and a
lightweight commercial model returned unusable output on every
fixture. Both exclusions predate the protocol freeze, and neither run
enters any analysis.

\FloatBarrier
\section{\mmcode{} extension detail}
\label{app:minimax}
The \mmcode{} campaign was added by a pre-data amendment to the
protocol, committed before any confirmatory response from this model
existed. The amendment fixes the same configuration, fixtures, and
$R=3$ as the primary campaigns, and defines an extension family with
its own three-hypothesis Holm adjustment, reported separately from the
two-model primary family throughout the paper. A single scout run of
this model predates the amendment. It is disclosed in the amendment
and is excluded from every confirmatory calculation.
Table~\ref{tab:minimaxext} lists the full extension results,
Table~\ref{tab:perrepmm} the per-replication counts and per-policy
invalid rates, and Table~\ref{tab:perclassmm} the per-class counts
corresponding to Table~\ref{tab:perclass} (descriptive only).

\begin{table}[H]
\centering
\begin{tabular}{lrrr}
\toprule
endpoint & RD (pp) & 95\% CI & $p$ \\
\midrule
E2 (\texttt{gr} $-$ \texttt{ss}) & +11.5 & [+5.9, +17.1] & 0.0002 \\
E1 (\texttt{vf} $-$ \texttt{gr}) & +2.8 & [-2.0, +7.5] & 0.2590 \\
E4 (portfolio $-$ \texttt{best\_of\_n}) & -9.0 & [-15.2, -2.7] & 0.0094 \\
E5 (\texttt{vf} $-$ \texttt{vf\_masked}) & -0.7 & [-5.3, +3.9] & 0.770$^\dagger$ \\
E3 (\texttt{vf}@1 $-$ \texttt{gr}@1) & +1.4 & [-2.0, +4.7] & -- \\
\bottomrule
\end{tabular}

\caption{Extension-family GEE endpoints for \mmcode{} (risk
differences in percentage points, fixture-clustered robust 95\% CIs).
E2, E1, and E4 carry the extension's own three-hypothesis Holm
adjustment. $\dagger$: unadjusted $p$ (E5 secondary). E3 is descriptive and
reported with effect and interval only.}
\label{tab:minimaxext}
\end{table}

\begin{table}[H]
\centering
\begin{tabular}{lcccc}
\toprule
policy & rep 1 & rep 2 & rep 3 & invalid \\
\midrule
\texttt{ss} & 18/145 & 19/145 & 14/145 & 11\% \\
\texttt{gr} & 30/145 & 33/145 & 38/145 & 38\% \\
\texttt{vf} & 38/145 & 36/145 & 39/145 & 45\% \\
\texttt{vf\_masked} & 34/145 & 48/145 & 34/145 & 34\% \\
\texttt{best\_of\_n} & 82/145 & 80/145 & 73/145 & 0\% \\
\bottomrule
\end{tabular}

\caption{Extension per-replication verifier-certified repair counts
(out of 145) and per-policy invalid rates.}
\label{tab:perrepmm}
\end{table}

\begin{table}[H]
\centering
\begin{tabular}{lcc}
\toprule
 & \multicolumn{2}{c}{\textsc{MiniMax-M2.5}} \\
class & \texttt{gr} & \texttt{vf} \\
\midrule
\texttt{drop\_w\_term} & 19/114 & 29/114 \\
\texttt{flip\_w\_sign} & 11/39 & 14/39 \\
\texttt{r\_charge\_perturb} & 41/156 & 44/156 \\
\texttt{rank\_perturb} & 30/126 & 26/126 \\
\bottomrule
\end{tabular}

\caption{Extension per-class verifier-certified repair counts, summed
over three replications. Descriptive only.}
\label{tab:perclassmm}
\end{table}

One provider constraint is disclosed in the amendment: the serving
platform locks this model's reasoning mode on and rejects requests
that try to disable it, so all extension attempts are generated with
provider-forced reasoning. Forced tool choice is rejected under this
mode, and the harness falls back to automatic tool choice, a
documented code path.

During the extension campaign, 45 records containing exogenous API
transport errors (connection failures and timeouts recorded by the
client) were quarantined and replaced by the filter-and-resume
machinery of the contamination rule, in an audited procedure with an
outcome-independent, presence-based selection predicate. The 45
records cover 43 distinct fixture-replications: one
\texttt{best\_of\_n} fixture was replaced again in two successive
passes after its rerun hit new transport errors. The removed records
are preserved byte-exactly, with a SHA-256 ledger, alongside the code.
Two disclosure sentences were fixed during the audit and are repeated
verbatim. First: ``All MiniMax contamination-cleaning reruns,
including the third rerun of
\texttt{spp\_N2\_d1\_node2\_rank\_perturb\_00}, retained the
pre-data-recorded 300-second client timeout; rows containing exogenous
transport timeouts were quarantined and replaced under the unchanged
filter-and-resume rule.'' Second: ``For pass 3 and any subsequent
contamination-cleaning reruns, requests to
\texttt{dashscope.aliyuncs.com} bypassed the shell's local HTTP(S)
proxy via \texttt{NO\_PROXY}, while the frozen 300-second client
timeout, provider endpoint, model, prompts, fixtures, policies, and
sampling parameters remained unchanged.''

\FloatBarrier
\section{Reproduction and configuration}
\label{app:repro}

The full pipeline is driven by five commands, all run inside the
repository's Python environment (\texttt{.venv}, created by the
installation instructions):
\begin{itemize}
\item \texttt{dualitycert generate-fixtures},
\item \texttt{dualitycert run-repair-loop}, one campaign per policy and
  replication, with \texttt{--arm} selecting the policy and
  \texttt{--model}, \texttt{--run-id}, and \texttt{--resume} for
  idempotent restarts,
\item \texttt{dualitycert score-e4},
\item \texttt{.venv/bin/python scripts/run\_gee.py} for the frozen GEE
  analyses, and
\item \texttt{.venv/bin/python scripts/paper\_tables.py} for
  every generated table and the results figure.
\end{itemize}

\begin{table}[H]
\centering
\begin{tabular}{lrrr}
\toprule
model & model calls & input tokens & output tokens \\
\midrule
\textsc{deepseek-chat} & 9,139 & 17.8M & 6.6M \\
\textsc{qwen-plus} & 8,944 & 14.7M & 6.6M \\
\bottomrule
\end{tabular}

\caption{Confirmatory-campaign compute accounting for the primary
models (all calls including failures).}
\label{tab:cost}
\end{table}

The
released repository pins the analysis-protocol freeze commit
(\texttt{813fdb0}), the tested harness commit (\texttt{26c5241}),
the SHA-256 of the fixture manifest, and the SHA-256 of every
configuration file in the execution manifest.

Sampling parameters were fixed before the confirmatory phase:
temperature at the provider default, a token cap of 8192 per call, a
client timeout of 300 seconds per attempt, five SDK retries with
exponential backoff, and forced tool choice, with the documented
automatic fallback for the extension model (Appendix~\ref{app:minimax}).
Table~\ref{tab:cost} accounts for the primary campaigns. The final
cleaned extension artifacts total 7{,}047 model calls with 13.0M input
and 24.0M output tokens. Including the 45 replaced records, the full
extension history is 7{,}324 calls with 13.4M input and 24.7M output
tokens.

The contamination audit trail ships with the code: a quarantine
directory with byte-exact prefilter copies of every affected file (the
first-pass removed rows are contained in these copies, and the later
passes also store their removed rows as separate files), the selection
predicate, a ledger of the affected fixture identities, and per-file
SHA-256 hashes for all nineteen quarantined record files, together
with per-run resume-segment sidecars and the full audit transcript.

\FloatBarrier


\end{document}